\documentclass[preprint,epsfig,floats,aps]{revtex4}
\usepackage{epsfig}
\begin{document}
%\baselinestretch{5}
%\tightenlines
\topmargin=-0.3cm
\title{Influence of the Pauli principle on the dynamics of the Quark-Gluon 
       Plasma}
%\vspace{0.1in}
\author{Ben-Hao Sa$^{1-3}$ and A. Bonasera$^2$ 
 \footnote{Email: sabh@iris.ciae.ac.cn; bonasera@lns.infn.it}}
\affiliation{
$^1$  China Institute of Atomic Energy, P. O. Box 275 (18),
      Beijing, 102413 China \\
$^2$  Laboratorio Nazionale del Sud, Istituto Nazionale Di Fisica Nucleare,
      Via S. Sofia 44, I-95123 Catania, Italy \\
$^3$  Institute of Theoretical Physics, Academy Sciences, Beijing,
      100080 China
}
%\maketitle
\begin{abstract}
A simple parton cascade model for the ultra-relativistic nucleus-nucleus 
collisions is proposed to investigate the Pauli blocking at the transition 
from hadronic to partonic state, i.e. QGP. A boost invariant study of the 
Pauli blocking is implemented in the Monte Carlo simulation for the first 
time. It turns out that the higher reaction energy the stronger the partonic 
Pauli effect is. It amounts to about ten percent at SPS and RHIC energies but 
nearly reaches twenty percent at LHC energy. Thus the effect is relevant to 
the formation and evolution of the quark-gluon plasma.
\end{abstract}
%\vspace{0.1in}
\maketitle

\begin{figure}[ht]
\centerline{\hspace{-0.5in}
\epsfig{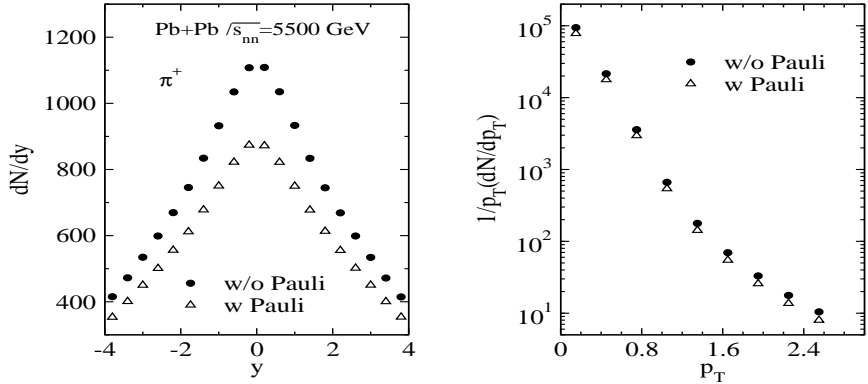}}
\vspace{0.2in}
\caption{Rapidity (left panel) and transverse momentum (right panel)
 distributions of $\pi^+$ in $Pb+Pb$ collisions at $\sqrt{s_{nn}}$=5500 GeV.}
\label{pau_pi}
\end{figure}

Pauli blocking plays an important role in nucleus-nucleus collisions at low 
and intermediate energies because the available phase space volume is quite 
small \cite{cug,bert,aich,aldo,sa1}. However, in the ultra-relativistic 
nucleus-nucleus collisions since the available momentum phase space is huge 
and more than 95$\%$ produced particles are bosons one might despise the role 
of Pauli effect. Even if the available momentum space in ultra-relativistic 
nucleus-nucleus collision is really huge, the coordinate space, on the 
other hand, is strongly contracted at the earliest stage of the collision. 
Although pions, kaons etc. do not suffer with Pauli effect, their precursors, 
u, d, and s quarks and their antiquarks, do really have all the influences 
of Pauli blocking if a Quark-Gluon Plasma (QGP) is formed and survives long 
enough in a limited region of phase space. Furthermore, since Pauli blocking 
is stronger for u and d quarks than s quark, it might benefit the production 
of strangeness relatively and the strong constrain of strange suppression 
could be somewhat overturned. Thus it is worth to study the effect of 
partonic Pauli blocking on the hadronic final state in ultra-relativistic 
nucleus-nucleus collisions.

\begin{figure}[ht]
\centerline{\hspace{-0.5in}
\epsfig{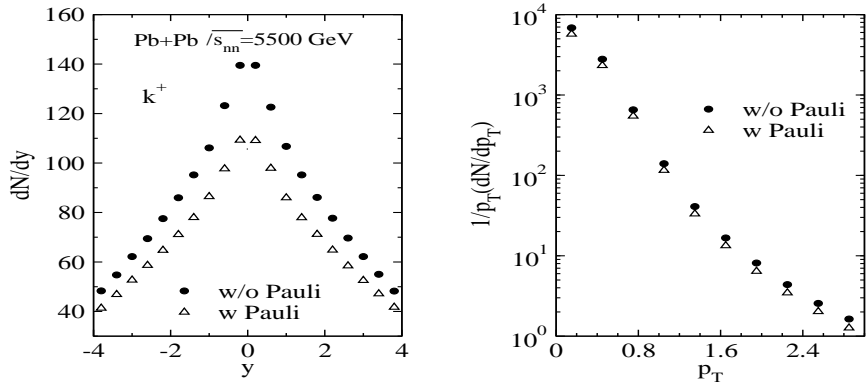}}
\vspace{0.2in}
\caption{Rapidity (left panel) and transverse momentum (right panel)
distributions of $k^+$ in $Pb+Pb$ collisions at $\sqrt{s_{nn}}$=5500 GeV.}
\label{pau_kp}
\end{figure}

Even though our parton cascade model (see below) is crude in some aspects, 
the basic assumption is that at each nucleon-nucleon collision the produced 
particles (mainly pions) are deconfined at least for the time of the 
reaction, a few fm/c for instance, which is however very short. If a QGP is
formed at the maximum compression stage we have then to test for the 
influence of Pauli principle since all the particles are fermions.
In our model the elementary probability of particle production are 
taken from data \cite{sjo1}, thus we expect the model to be reasonable in 
describing the dynamics of particle production. However, the mean field 
effects and the dynamics of confinement are neglected and we assume that the 
quarks recombine simply when the density in phase space is small. The later, 
however, is a rather reasonable assumption since if the density is very high, 
the quarks are clearly deconfined. As we will show below, our calculations 
underestimate the available experimental data nearly a factor of 1.5, thus we 
expect that we are underestimating the Pauli blocking since it is 
more severe for larger densities (i.e. higher number of produced partons). 
More refined models which include the dynamics of the QGP plasma must really 
take the Pauli blocking into account. Simpler models which deal with hadrons 
only, should estimate the maximum density in phase space reached. If such 
density is too large then it is necessary to deal with the structure of the 
hadrons. From an experimental point of view, since the quarks recombine 
mainly in pions which are bosons, most of the Pauli blocking effect might be 
hidden. However, some important informations on the phase space density might 
be obtained from baryon-baryon coincidence measurement, especially neutrals 
to avoid the distortions of the long range Coulomb force.

\begin{figure}[ht]
\centerline{\hspace{-0.5in}
\epsfig{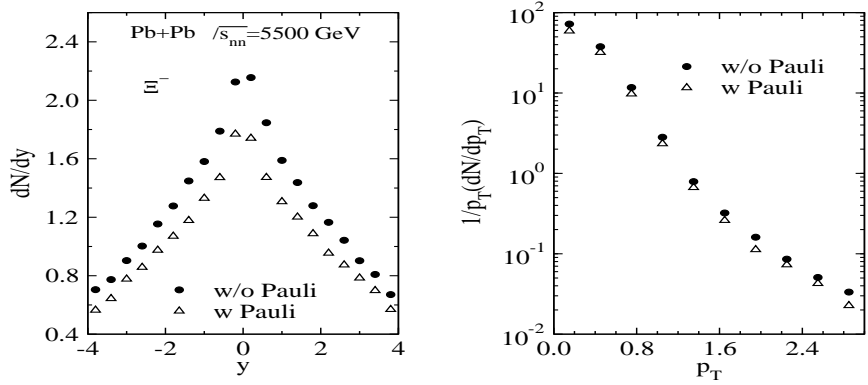}}
\vspace{0.2in}
\caption{Rapidity (left panel) and transverse momentum (right panel)
 distributions of $\Xi^-$ in $Pb+Pb$ collisions at $\sqrt{s_{nn}}$=5500 GeV.}
\label{pau_ca}
\end{figure}

To our knowledge the quantum statistic effect in ultra-relativistic heavy 
ion collisions was really implemented in the study of bosonic enhancement 
of pion transverse momentum \cite{bert1} and was also considered formally in 
the partonic transport equation \cite{gei}. It should be emphasized that 
a great effort was made in \cite{gei} after the pioneering study \cite{boa} 
in the field of parton cascade model. However, the partonic Pauli blocking 
effect was not really incorporated in the Monte Carlo simulation of currently 
active parton cascade models \cite{bin,gyu,boer,bass,zhe}. In this letter a 
simple parton cascade model for ultra-relativistic nucleus-nucleus collision 
is proposed investigating the effect of Pauli blocking at partonic 
(initialization) stage on the hadronic final state. The partonic Pauli 
blocking is implemented boost invariantly for the first time in Monte Carlo 
simulation. The amount of this Pauli effect is below ten percent at SPS and 
RHIC energies but reaching nearly twenty percent at LHC energy, due to the 
increased number of produced partons with increasing beam energy. 

\begin{table}[hb]
\tiny{
\begin{tabular}{c|c|c|c|c|c|c|c|c|c|c|c|c}
\multicolumn{13}{c}{Table 1. Particle yield in $Au+Au$ collisions at $\sqrt{s_
{nn}}$=200 and 130 GeV}\\
\multicolumn{13}{c}{and $Pb+Pb$ collisions at $\sqrt{s_{nn}}$=5500 and 17.3 
GeV$^1$}\\
\hline\hline
 &\multicolumn{3}{c|}{$Pb+Pb$} &\multicolumn{3}{c|}{$Au+Au$}& \multicolumn{3}
{c|}{$Au+Au$}& \multicolumn{3}{c}{$Pb+Pb$} \\
\cline{2-13}
 &\multicolumn{3}{c|}{$\sqrt{s_{nn}}$=5500 GeV} &\multicolumn{3}{c|}{$\sqrt
{s_{nn}}$=200 GeV} &\multicolumn{3}{c|}{$\sqrt{s_{nn}}$=130 GeV} 
&\multicolumn{3}{c}{$\sqrt{s_{nn}}$=17.3 GeV} \\
\cline{2-13}
 Particle &$w/o$ Pauli &$w$ Pauli &1-$\frac{w}{w/o}$ &$w/o$ Pauli &$w$ Pauli &
1-$\frac{w}{w/o}$ &$w/o$ Pauli &$w$ Pauli &1-$\frac{w}{w/o}$ &$w/o$ Pauli &
$w$ Pauli &1-$\frac{w}{w/o}$\\
\hline
 $\pi^+$ &7291 &6034 &0.172 &1352 &1261 &0.0673 &1114  &1058  &0.0503 &427.8 
&416.5 &0.0264 \\
 $\pi^-$ &7308 &6050 &0.172 &1367 &1276 &0.0666 &1129  &1073  &0.0496 &441.0 
&429.3 &0.0265 \\
 $k^+$   &856.0 &712.7 &0.167 &162.1 &151.7 &0.0642 &133.6 &127.2 &0.0490 
&52.54 &51.08 &0.0278 \\
 $k^-$ &819.1 &677.7 &0.173 &128.1 &118.5 &0.0749 &100.2 &94.31 &0.0588 
&26.28 &25.53 &0.0285 \\
 $k^0_s$ &838.2 &694.2 &0.172 &145.2 &135.2 &0.0689 &117.2 &111.1 &0.0520 
&39.64 &38.57 &0.0270 \\
 $\Lambda$ &168.0 &145.8 &0.132 &29.98 &28.66 &0.0440 &24.17 &23.43 &0.0306 
&7.037 &6.947 &0.0128 \\
 $\bar{\Lambda}$ &157.9 &132.1 &0.163 &24.06 &22.56 &0.0623 &18.93 &18.13 
&0.0423 &4.463 &4.343 &0.0269 \\
 $\Xi^-+\overline{\Xi^-}$ &23.52 &19.76 &0.160 &3.670 &3.469 &0.0548 &2.910 
&2.770 &0.0481 &0.6681 &0.6516 &0.0247 \\
 $\Omega^-+\overline{\Omega^-}$ &0.601 &0.503 &0.164 &0.0920 &0.08444 &0.0822 
&0.0747 &0.0691 &-0.0752 &0.0138 &0.0134 & -0.0347 \\
 strangeness$^2$ &2863 &2383 &0.168 &493.2 &459.3 &0.0675 &397.1 &377.0 
&0.0505 &130.6 &127.2 &0.0260 \\
\hline
\hline
\multicolumn{13}{l}{$^1$ no feed-down correction.} \\
\multicolumn{13}{l}{$^2$ The sum of the yield of strange particles.}
\end{tabular}
}
\end{table}

The parton cascade model for ultra-relativistic heavy ion collision is 
generally composed of the parton initial state, the parton evolution, the 
hadronization, and the hadron evolution. However, the models in 
\cite{boer,zhe} lack the last two parts at present, for instance. The 
partonic Pauli blocking, of course, plays a role in the former two parts 
only. As a first step the proposed simple parton cascade model is used for 
investigating mainly the Pauli blocking effect in the generation of parton 
initial state from the hadronic distributions at the highest density stage. 

There are two ways of creating the parton initial state. In 
\cite{bin,gyu,zhe} the parton initial state was composed of partons from the 
mini-jets production in nucleus-nucleus collision and the HIJING multiple 
mini-jet generator \cite{wang} was specified in \cite{bin,gyu}. The parton 
initial state in \cite{boer,bass} was created via probability distributions 
first for the spatial and momentum coordinates of nucleons in colliding 
nuclei and then for the flavor and spatial and momentum coordinates of 
partons in nucleons. Our parton cascade model follows the former way, 
however, the JPCIAE multiple mini-jet generator \cite{sa2} is used instead 
of HIJING.

The JPCIAE multiple mini-jet generator for ultra-relativistic nucleus-nucleus 
collision is based on PYTHIA \cite{sjo1} which is a well known event 
generator for hadron-hadron collision. In the JPCIAE model the radial 
position of a nucleon in colliding nucleus A (indicating the atomic number of 
this nucleus as well) is sampled randomly from Woods-Saxon distribution. Each 
nucleon is given a beam momentum in z direction and zero initial momenta in x 
and y directions (fermionic motion is neglected). The Lorentz contraction is 
taken into account after the initialization of nucleons. A least approaching 
distance for each colliding nucleon pair along their straight line trajectory 
is calculated together with its collision time under the requirement that the 
least approaching distance must be less than or equal to $\displaystyle{\sqrt
{\sigma_{tot}/\pi}}$. Here $\sigma_{tot}$ refers to the total cross section 
assumed to be equal to 50 mb for $Au+Au$ collisions at $\sqrt{s_{nn}}$=130 
and 200 GeV and 40 and 100 mb for $Pb+Pb$ collisions at $\sqrt{s_{nn}}$=17.3 
and 5500 GeV \cite{pdg}, respectively. The nucleon-nucleon collision with the 
least collision time is then selected from the initial collision (time) list 
performing the first collision. This nucleon-nucleon collision is modeled by 
PYTHIA with string fragmentation switched off, thus the produced particles 
are quark pairs, diquark pairs and gluons. To investigate the partonic Pauli 
blocking efficiently the diquark (anti-diquark) is split into quarks 
(antiquarks) and the gluons into quark pair randomly. The produced partons 
propagate along straight line trajectory similarly to the nucleons. However, 
the partonic interaction is neglected for the moment, thus partons do not 
collide and possibly create new partons, which is a possible reason for 
underestimating the data. After a nucleon-nucleon collision both the 
particle list and the collision list are updated and the new collision list 
is still composed of nucleon-nucleon collisions only. The next collision is 
selected from the new collision list and the processes above are repeated 
until the collision list is empty. 

For each parton $i$ among $N_{new}$ partons produced in the current nucleon
-nucleon collision a judgment must be performed for its Pauli blocking and 
its unblocking probability, $p_{unb}(i)$, is calculated (see below). If the 
product of unblocking probabilities of $N_{new}$ produced partons 
\begin{equation}
P_{unb}=\prod_{i=1}^{N_{new}}p_{unb}(i)
\end{equation}  
satisfies 
\begin{equation}
P_{unb} \leq \xi
\end{equation} 
where $\xi$ is a random number, the current nucleon-nucleon collision is 
blocked and thrown away. Another nucleon-nucleon collision is selected from 
the collision list and the processes above are repeated.  

In the heavy-ion collision at low and intermediate energies the Pauli effect 
was treated boost uninvariantly \cite{cug,bert,aich,aldo,sa1}. We introduce 
a method of boost invariant Pauli blocking in Monte Carlo simulation for 
the first time. To calculate $p_{unb}(i)$ we first select the partons with 
same flavor as parton $i$ from both the $N_{new}$ partons and the particle 
list formed before current nucleon-nucleon collision. If the total number of 
partons picked above is denoted by $N_{pick}$ we boost all of $N_{pick}$ 
partons to the rest frame of parton $i$ to be consistent with the method of 
least approaching distance (this boost is not necessary indeed, we have 
checked that the discrepancy between results from with and without this 
boost is nearly 1\%.). Two three dimensional cubes: one in coordinate space 
with size $\Delta_r$ and the other in momentum space with size $\Delta_p$, 
are defined at $i$ (as origin). The product of these two cubes, which is an 
invariant scalar \cite{las}, is assumed to be $h^3$, the consequent values 
of $\Delta_r$ and $\Delta_p$ are 1 fm and 1.24 GeV/c, respectively. If the 
six dimensional cell spanned between partons $i$ and $j$ 
($j$ $\epsilon$ $N_{pick}$), $|{x_{ij}\times y_{ij}\times z_{ij}\times px_j
\times py_j\times pz_j}|$ (invariant scalar), is within $(\Delta_r \times 
\Delta_p)^3$, i. e. the conditions  
\begin{eqnarray}
|x_{ij}|\leq\Delta_r/2, |y_{ij}|\leq\Delta_r/2, |z_{ij}|\leq\Delta_r/2,\\
|px_j|\leq\Delta_p/2, |py_j|\leq\Delta_p/2, |pz_j|\leq\Delta_p/2,
\end{eqnarray}
are satisfied simultaneously, the occupation number of parton with flavor as 
parton $i$, $N_{occu}(i)$, is added by one. The $x_{ij}$ and $px_j$ in  
the above equations, for instance, are the $x$ component of the coordinate 
distance between $i$ and $j$ and of the momentum of $j$ in rest frame of $i$ 
($i$ $\epsilon$ $N_{new}$), respectively. Thus the occupation probability of 
parton with flavor as parton $i$ reads 
\begin{equation}
p_{occu}(i)=N_{occu}(i)/g
\end{equation}   
where $g$=6 is the spin and color degeneracies of the quark and/or antiquark 
with a given flavor (identified as u, d, and s). The unblocking probability 
of parton $i$ is then 
\begin{equation}
p_{unb}(i)=1-p_{occu}(i). 
\end{equation} 

The above partonic system is then hadronized by JETSET \cite{sjo1} after the  
nucleon-nucleon collision ceased (freeze-out) and the hadronic final state 
is the consequence. One investigates the partonic Pauli blocking in the 
generation of the initial parton state and consequently on the hadronic 
final state by comparing the calculations with partonic Pauli blocking to 
those without.  

In Tab. 1 is given the multiplicity of particles in $Au+Au$ collisions at 
$\sqrt{s_{nn}}$=130 and 200 GeV (5$\%$ most central) and $Pb+Pb$ collisions 
at 17.3 GeV (5$\%$ most central) and 5500 GeV (10$\%$ most central) from 
calculations with and without Pauli blocking. The decreasing percentage of 
the particle multiplicity in the calculations with Pauli blocking relative 
to the ones without Pauli blocking is given in the table as well. In the last 
row the strangeness multiplicity, the sum of all the strange particles in the 
table, is shown. One sees in Tab. 1 that the partonic Pauli blocking effect 
is generally below ten percent at SPS and RHIC energies but reaching nearly 
twenty percent at LHC energy. The higher reaction energy the stronger Pauli 
effect because of the increasing number of produced quarks and of the 
competition between the partonic Pauli blocking, which decreases pion 
production, and the fragmentation, which benefits low $p_T$ pions. The 
partonic Pauli effect is also of benefit to the production of strangeness as 
seen in $Pb+Pb$ collision at LHC energy.  

The rapidity (left panel) and transverse momentum (right panel) distributions 
with and without Pauli blocking for $Pb+Pb$ collisions at $\sqrt{s_{nn}}$=
5500 GeV are given, respectively, in Fig. \ref{pau_pi}, \ref{pau_kp}, and 
\ref{pau_ca} for $\pi^+$, $k^+$ and $\Xi^-$, for instance. The discrepancy 
between the two cases is visible, in rapidity distribution especially.  

In summary, a simple parton cascade model for ultra-relativistic nucleus
-nucleus collisions is proposed studying the partonic Pauli blocking effect 
in the parton initial state and consequent hadronic final state. A boost 
invariant study of the Pauli effect is implemented in the Monte Carlo 
simulation for the first time. It turned out that the amount of this Pauli 
blocking effect is below ten percent at SPS and RHIC energies but reaching 
nearly twenty percent at LHC energy, i.e. the higher beam energy the stronger 
the partonic Pauli effect is. Thus the effect is relevant to the formation 
and evolution of the quark-gluon plasma. However, the results from our model 
are a kind of under estimation. Since the charge multiplicity in $Au+Au$ 
collisions at $\sqrt{s_{nn}}$=200 and 130 GeV in the calculations with Pauli 
blocking is, respectively, 3063 and 2562 (including feed-down correction from 
$k^0_s$, $\Lambda$, and $\bar{\Lambda}$) and the multiplicity of $\pi^+$ and 
$\pi^-$ in $Pb+Pb$ collision at $\sqrt{s_{nn}}$=17.3 GeV is, respectively, 
416.5 and 429.3 in Tab. 1. They are all less than the corresponding 
experimental data of 4630$\pm$370 \cite{bear}, 3860$\pm$300 \cite{bear1}, and 
619$\pm$17$\pm$31 and 639$\pm$17$\pm$31 \cite{afa}, respectively. On the 
other hand, the results of our parton cascade model are not too bad to 
compare with experimental data, which means that the model is reasonable in 
its range.      

Finally, the financial support from NSFC (10135030, and 10075035) in China 
and INFN and Department of Phys. University of Catania in Italy (where most 
of the work was performed) are acknowledged.

\end{document}